\documentclass[aip,jcp,preprint,unsortedaddress,a4paper,onecolum]{revtex4-1}

\usepackage[fleqn]{amsmath}
\usepackage{amssymb}
\usepackage[dvips]{graphicx}
\usepackage{color}
\usepackage{tabularx}
\usepackage{algorithm}
\usepackage{algorithmic}


\begin{document}

\title{Adaptive Resolution Molecular Dynamics Technique: Down to the Essential}
\author{Christian Krekeler}
\affiliation{Institute for Mathematics, Freie Universit\"at Berlin, Germany}
\author{Animesh Agarwal}
\affiliation{Theoretical Biology and Biophysics Group, Los Alamos National Laboratory, Los Alamos, NM 87545, USA}
\author{Christoph Junghans}
\affiliation{Computer, Computational, and Statistical Sciences Division, Los Alamos National Laboratory, Los Alamos, NM 87545, USA}
\author{Matej Praprotnik}
\affiliation{Laboratory for Molecular Modeling, National Institute of Chemistry, Hajdrihova 19, SI-1001 Ljubljana, Slovenia and Department of Physics, Faculty of Mathematics and Physics, University of Ljubljana, Jadranska 19, SI-1000 Ljubljana, Slovenia}
\author{Luigi Delle Site}
\affiliation{Institute for Mathematics, Freie Universit\"at Berlin, Germany}
\email{luigi.dellesite@fu-berlin.de}
\begin{abstract}
  We investigate the role of the thermodynamic (TD) force, as an essential
  and sufficient technical ingredient for an efficient and accurate adaptive resolution algorithm. Such a force  applied in the coupling region
  of an adaptive resolution Molecular Dynamics (MD) set-up, assures thermodynamic equilibrium between atomistically resolved and coarse-grained regions,
  allowing the proper exchange of molecules. We numerically prove that indeed for systems as relevant as liquid water and 1,3-dimethylimidazolium chloride ionic liquid, the
combined action of the TD force and thermostat allows for computationally efficient and numerically accurate simulations, beyond the current
capabilities of adaptive resolution set-ups, which employ switching functions in the
coupling region.
\end{abstract}

\maketitle
\section{introduction}
\begin{figure}[htbp]
\centering
\includegraphics[clip=true,trim=0.1cm 0cm 0cm 0.1cm,width=13cm]{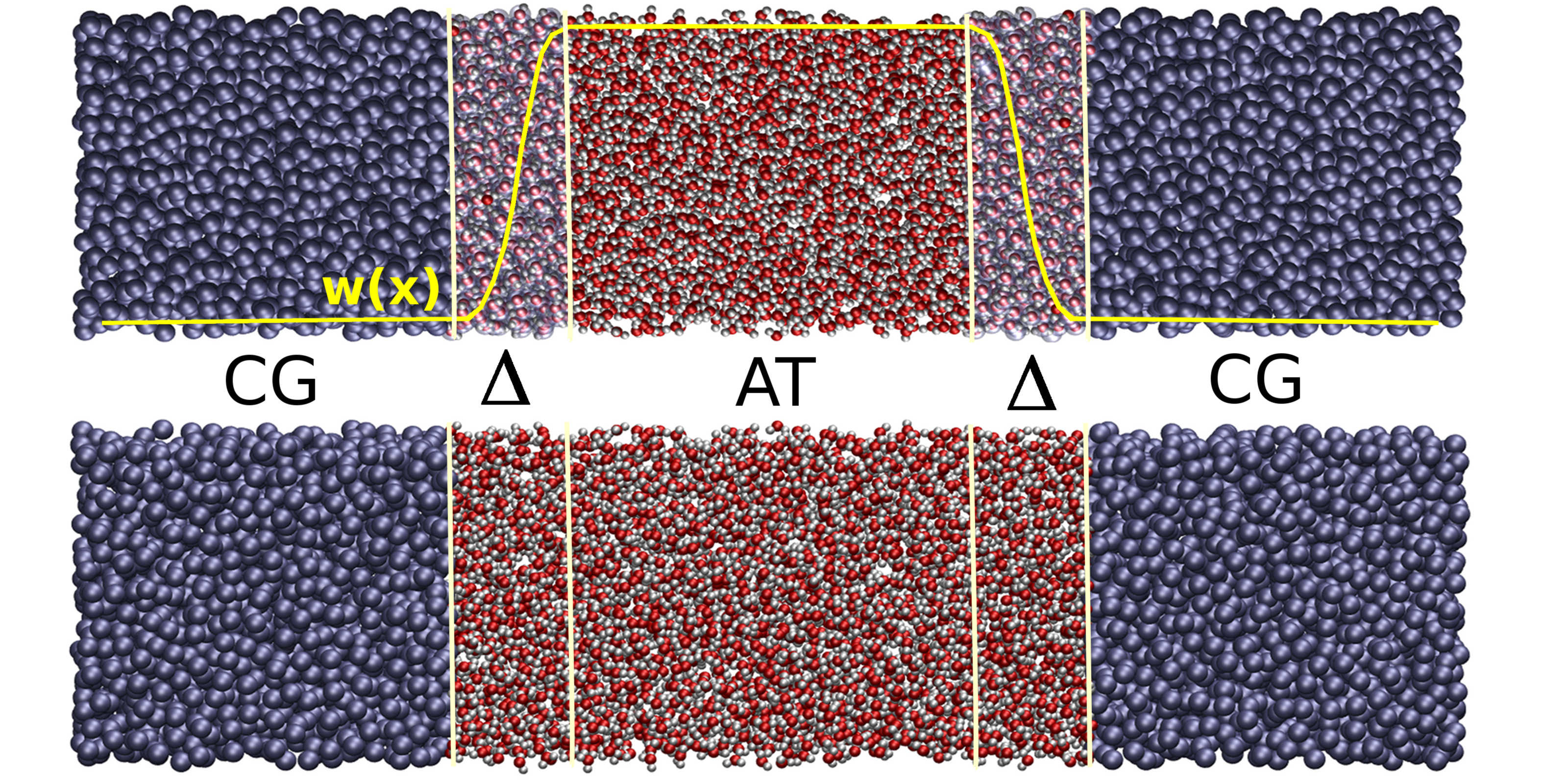}
\caption{(Top) Set-up of standard AdResS: in the central AT region
  molecules have atomistic resolution. The AT region is interfaced with the
  hybrid region $\Delta$, where molecule have hybrid atomistic/coarse-grained
  resolution weighted by the switching function $w(x)$, illustrated in
  yellow. $\Delta$ is interfaced with the coarse-grained region CG.
(Bottom) Set-up investigated in the current work. There are only two
  regions with different resolution: atomistic and coarse-grained. In
  $\Delta$, atomistic molecules interact with coarse-grained molecules via
  the coarse-grained potential. In this sense, they have an effective hybrid
  resolution and thus $\Delta$ is equivalent to the $\Delta$ of the top
  cartoon without spatial interpolation.}
\label{cartoon}
\end{figure}

Recent developments in the field of multiscale MD methods have brought the idea of adaptive molecular resolution for simulation of liquids to a high standard level, both conceptually and technically \cite{jcp-adress,annurev, ensing1,truhlar,csany}.
In particular, among all methods, the Adaptive Resolution Simulation (AdResS)
\cite{jcp-adress,annurev,pre1-adress} has been intensely developed
and applied to a large number of systems in different fields (see recent
Refs. \cite{matej-shear,matej-dna,kurt-nucl-ac,anim-pccp,shad-adts} as an example of
variety). The root model of the adaptive resolution technique (AdResS) is based on partitioning the
simulation box in three regions, namely: a region at atomistic resolution (AT
region), a region at coarse-grained resolution (CG region) and, as an
interface between AT and CG, a hybrid region ($\Delta$ region) (see
Fig.~\ref{cartoon}). 
In the $\Delta$ region, the force between molecules $\alpha$ and $\beta$ is defined by a space-dependent interpolation of the AT and CG forces: $F_{\alpha \beta} = w(X_{\alpha})w(X_{\beta})F_{\alpha \beta}^{AT} + [1- w(X_{\alpha})w(X_{\alpha})]F_{\alpha \beta}^{CG}$; $w(x)$ is the interpolating function, smoothly passing from the value 1 in AT to the value 0 in CG.
Although based on empirical arguments, such an algorithm has been proven to be computationally robust for several highly challenging applications (see e.g. \cite{wat1,wat2,full1,prlado}). Further developments of the root model have removed most of the empirical character and led AdResS through two different but complementary routes: (a) Grand Canonical MD for the AT region (GC-AdResS) and Open Boundary MD (OBMD) for the AT region and coupling to continuum \cite{prl2012,prx,njp,obmd,obmd-DNA}; (b) Global Hamiltonian for the entire simulation box (H-AdResS) \cite{prlraff1,prlraff2}. The former consists of re-framing the force-based technique within a solid conceptual background of Grand Canonical ensemble for AT, with mathematically sound criteria of validity (see in particular Refs.\cite{prx,njp}). The latter instead introduces a technical variation, that is the space-dependent coupling of the potential instead of forces: $U_{\alpha \beta} = \lambda(\hat{X}_{\alpha\beta})U_{\alpha \beta}^{AT} + [1- \lambda(\hat{X}_{\alpha\beta})]U_{\alpha \beta}^{CG}$; $\lambda(\hat{X}_{\alpha\beta})$ is equivalent to $w(x)$ but acting on $\hat{X}$, the center of mass of molecules $\alpha$ and $\beta$. Common to both derivations is the addition of a one-body external thermodynamic term that, while conceptually derived from basic principles of statistical mechanics, in technical terms allows for a density balance between the various regions. In GC-AdResS (and AdResS), this term is named a TD force and is calculated self-consistently in the equilibration step before the production run: $F_{k+1}^{TD}(x)=F_{k}^{TD}(x) - \frac{M}{[\rho_{ref}]^2\kappa}\nabla\rho_{k}(x)$; $M$ is the mass of the molecule, $\kappa$ a (converge-driven, but well defined) tunable constant, $\rho_{k}(x)$ is the molecular density  as a function of the position in the $\Delta$ region at the $k$-th iteration \cite{prl2012}. Conceptually, its application corresponds to equalizing the chemical potential of the various resolutions to the reference atomistic chemical potential \cite{jcp-simon,jctc-han,prx,jcp-mu}. Instead in H-AdResS, the density balancing operation is made in terms of a balance of an artificial global free energy (because a global Hamiltonian with a physical meaning cannot exist \cite{pre2007}), in an adiabatic interpretation of the switching from the AT to the CG potential in the $\Delta$ region (Kirkwood Hamiltonian)\cite{prlraff1,jcppep}. Independently from the specific coupling, both routes go under the guiding principles for systems with open boundaries \cite{physrep}.
Recent applications of GC-AdResS for extremely challenging systems as imidazolium-based ionic
liquids, i.e., a high density liquid mixture of molecular anions and cations, have highlighted the capability  of the TD force in establishing
equilibrium in the system and produce results of high standard
\cite{krek,shadrack,krekpshad}. As a consequence, a natural question concerns the
possibility that, contrary to the pioneering idea of AdResS, the
TD force would be actually a more fundamental technical characteristics for the
adaptivity than the smooth interpolation of the $\Delta$ region. Additional
arguments to support the concept that the TD force, as formulated
above, has a fundamental role in equilibrating and stabilizing a system, are
provided also in recent work of other groups (see, e.g., Ref. \cite{hmmm}).


In this work, we investigate the possibility that the TD force represents the
essential technical ingredient in the molecular resolution adaptation process and show that indeed it
is able to assure proper equilibrium and stability even in case of a direct
AT/CG interface without any space-dependent smoothing (see Fig.~\ref{cartoon}). In such a case, standard
AdResS becomes automatically H-AdResS with $\lambda(\hat{X}_{\alpha\beta})=1$ and vice versa.
Despite that such a scheme implies the existence of a global Hamiltonian,
our interest will not be related to this aspect. As reported in
\cite{physrep}, although in MD Hamiltonian-based algorithms represent a traditional source of genuine rigor, this principle does not necessary hold anymore for systems with open boundaries that exchange matter with an environment.
Actually, for AdResS, at a conceptual level the existence of a global Hamiltonian does not imply a
mathematical superiority compared to a force-based approach; on the contrary, as anticipated above, a global
Hamiltonian implies a unphysical global (micro-)canonical ensemble which may lead to a misinterpretation of results (see also \cite{pre2007}). Moreover, in actual adaptive resolution simulations of realistic systems, energy dissipation is, for several reasons, inevitable even if one has a well defined global Hamiltonian. For example, it is mandatory to apply a repulsive force capping that prevents atoms overlapping in the $\Delta$ region even for molecules of medium size in the standard interpolation scheme. Such a force capping induces dissipation and requires the action of a thermostat for achieving the desired chemico-physical equilibrium \cite{jan}. Furthermore, an equilibrating thermostat is mandatory for performing simulations (even full-blown atomistic ones) of any system with a finite cut-off for the interactions, e.g., the reaction field method\cite{neumann} for electrostatics used in AdResS. It must be also underlined that the hybrid or transition region where the two resolutions are coupled is an artificial region by construction, in any AdResS version. For this reason, one is free to operate at technical level in such regions within some generic/macroscopic constraints (see discussion below) to make the algorithm more efficient. The key point of such an approach is that despite the artificial character of the $\Delta$ region, physical properties of the AT region do not depend on the pragmatic computational (empirical) parametrization of the transition region. In fact in Ref.\cite{njp} AdResS was mapped onto the Bergman-Lebowitz model of open system with the only constraint that the $\Delta$ and CG region are, at any time, at the same molecular density and temperature of the thermodynamic point chosen for the AT region. In this sense, (GC-)AdResS for the atomistic region, considered as an open region embedded in a generic thermodynamic environment, is not an empirical approach but has got a real physical meaning at any time.
In this perspective, here we will focus on the technical validity of the method in reproducing, in the AT region, results of a reference NVT full-blown atomistic
simulation, that is, as underlined above, we will consider the AT region as a system with open
boundaries (Grand Ensemble) \cite{njp,physrep}. A major technical advantage of this formulation concerns the simplicity of implementation and transferability of the algorithm in any MD code. Its implementation consists only of (i) the partitioning of the simulation box in two regions, (ii) the definition of the CG molecules in the neighboring list of the AT molecules at the border (and vice versa), (iii) the addition of the external iterative loop for the convergence of the TD force, and (iv) if necessary, finding a proper molecular orientation avoiding molecular overlaps when entering the $\Delta$ region from the CG side (see below). A side consequence of such simplicity is the removal of internal loops required by the switching function, $w(X_{\alpha})w(X_{\beta})$ (or $\lambda(\hat{X}_{\alpha\beta})$ in H-AdResS), in locating a specific pair of molecules at each time step. Such a removal leads to a sizable gain in computational performances, as it was predicted by a previous computational analysis of AdResS \cite{amdahl}.
There is also one relevant methodological advantage, namely that the atomistic
radial distribution functions (RDFs) are the same as in a full atomistic simulation
not only in the AT region but also in the $\Delta$ region. This condition automatically
satisfies the necessary condition derived in \cite{prx} which assures, at mathematical/rigorous
level, that the probability distribution function in the AT region is the same
of a full atomistic simulation up to a third order (i.e. up to a 3-particle
distribution function); this property in MD implies a very high accuracy (within $2-3\%$) in the statistical determination of physical properties in the AT region. Next, we
will report the formal structure of the direct coupling  and discuss its
numerical implementation. We will then show numerical results for liquid water at ambient conditions; in addition, with the current method, we will reproduce some relevant data of a recent published work regarding the solvation and the free energy of aggregation of two micelles in water. The original results were obtained with the previous approach at a higher computational cost compared to the computational effort required by the current approach. Finally, we will show numerical results for 1,3-dimethylimidazolium chloride ionic liquid in the extreme case of uncharged coarse-grained models for the anions and cations.

\section{Methodology: From smooth coupling to a direct interface}
Following Fig.~\ref{cartoon}, the simulation
box is divided in an AT and a CG region. Let us define $d$, the
cut-off distance of molecular interactions. In the AT region, all
molecules interact via the atomistic potential:
$V^{AT}=\sum_{(\alpha,\beta)\in AT}V^{AT}({\bf r}_{\alpha}, {\bf r}_{\beta})$, with $V^{AT}({\bf r}_{\alpha}, {\bf r}_{\beta})$ the atomistic potential between all the atoms ${\bf r}_{\alpha}, {\bf r}_{\beta}$ of molecule $\alpha$ and molecule $\beta$.
Similarly in the CG region, all
molecules interact via a coarse-grained potential: $V_{CG}=\sum_{(\alpha,\beta)\in CG}V^{CG}({\bf R}_{\alpha}, {\bf R}_{\beta})$, with
$V^{CG}({\bf R}_{\alpha}, {\bf R}_{\beta})$ the coarse-grained potential between the centers of mass
${\bf R}_{\alpha}, {\bf  R}_{\beta}$ of molecule $\alpha$ and molecule $\beta$. In the AT region at the interface,
that is in the region of dimension $d$ along the $x-$axis, named region $\Delta$ in \ref{cartoon},
AT  molecules interact with CG molecules via the coarse-grained potential, i.e. the coupling AT-CG
interaction. Since CG molecules do not have
atomistic degrees of freedom there is no other way to interact
with AT molecules:
$V^{coupling}_{\Delta}=\sum_{\alpha\in\Delta}\sum_{\beta\in CG}V^{CG}({\bf R}_{\alpha},{\bf
  R}_{\beta})$, with ${\bf R}_{\alpha}$ the center of mass of the atomistic
molecule $\alpha$ and ${\bf R}_{\beta}$ the center of mass of the
coarse-grained molecule $\beta$. Effectively, the $\Delta$ region acts
similarly to the transition region of standard AdResS. Thus, atomistic
molecules closer to the GC region experience the CG character of the
interaction more than AT molecules located at larger distance from the CG region; that is the
passage from one resolution to the other is not artificially smoothed via
$w(x)$ as in standard AdResS, but it is implicitly gradual (i.e. function of
$x$, not strictly abrupt). Similarly to standard AdResS and following the same
numerical procedure, in $\Delta$,  we define the TD force for the
density balance. As already explained, one of the aims of the switching function $w$ in combination with the
repulsive force capping and the local thermostat in the standard AdResS is to avoid fatal large forces due to potential overlaps (or unphysical
short distances) between atoms of neighboring molecules at the CG/$\Delta$ interface.
In the new implementation, the overlaps must be avoided in another way, e.g., by fixing the center of mass of 
a problematic molecule and running a few MD steps with a repulsive force capping to find an energetically permissible orientation. In a Monte-Carlo implementation, for example, this would be automatically taken care of by the rejection criterion \cite{cameron}. In the present work, we adopt the simplest approach, that is we avoid molecular overlaps by simply capping the interactions at close distances. These are all the technical ingredients required for accurate and efficient GC-AdResS simulations of liquids. Conceptually (but not technically), this should be equivalent to the standard approach with the Heaviside step function used as the switching function $w$ supplemented with the molecular orientation adjustment.
\section{Results}
\subsection{Liquid water at room condition}
We consider systems of liquid water at room conditions of different sizes, as treated in
previous work with AdResS and whose technical details are reported
in the Appendix. Here, we report the results for a system of $7000$
molecules in total. Our aim is to show the accuracy of the scheme in
predicting basic properties in the AT region. The necessary condition that any
adaptive resolution technique must satisfy is that the particle number density
is uniform across the simulation box. Such a goal in AdResS is achieved
through the combined action of the TD force and a thermostat so
that the thermodynamic state point corresponds to the desired state of
reference. 
 \begin{figure}[htbp]
	        \centering
                \includegraphics[clip=true,trim=0cm 0cm 0cm
                0cm,width=10cm]{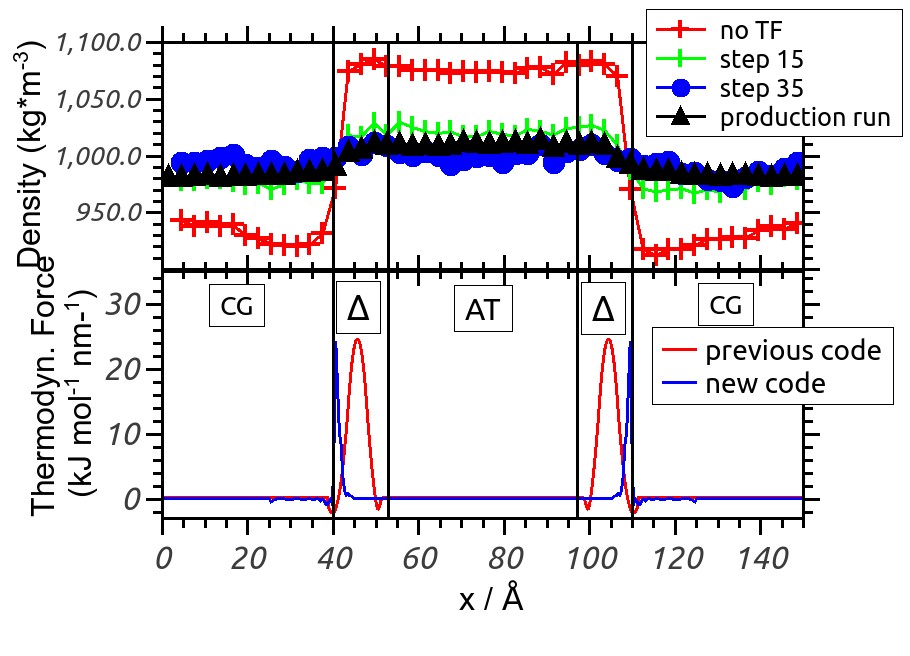}
                \caption{upper panel:The particle number density converges to the target
                value after the iterative application and calculation of the
                TD force. The curve in black represents the density
                during the production run. Lower panel: Thermodynamic force as a function of the 
                position in $\Delta$. In red, for the previous approach with a smoothed transition, 
                and in blue, for the current approach. The action of the force is sharply 
                concentrated around the direct interfacial region, as one would expect. 
                The extension of the force outside $\Delta$ is the AT and CG is due to a technical 
                convenience in the iterative calculation, however, the action in such regions is negligible.}
                \label{dens}
                \end{figure}
Fig.~\ref{dens} depicts the convergence of the density to the target
value  (within a $3\%$ accuracy) as the calculation of the TD force
progresses. Once the TD force is determined (see lower panel of Fig.\ref{dens}) the density is
maintained uniform for the production run.
Compared to the standard AdResS, the new
scheme requires, as expected, some more iteration steps, however, their cost is
negligible. Next, we must prove that structural properties are accurately
reproduced. 
\begin{figure}[htbp]
	        \centering
                \includegraphics[clip=true,trim=0cm 0cm 0cm
                0cm,width=6cm]{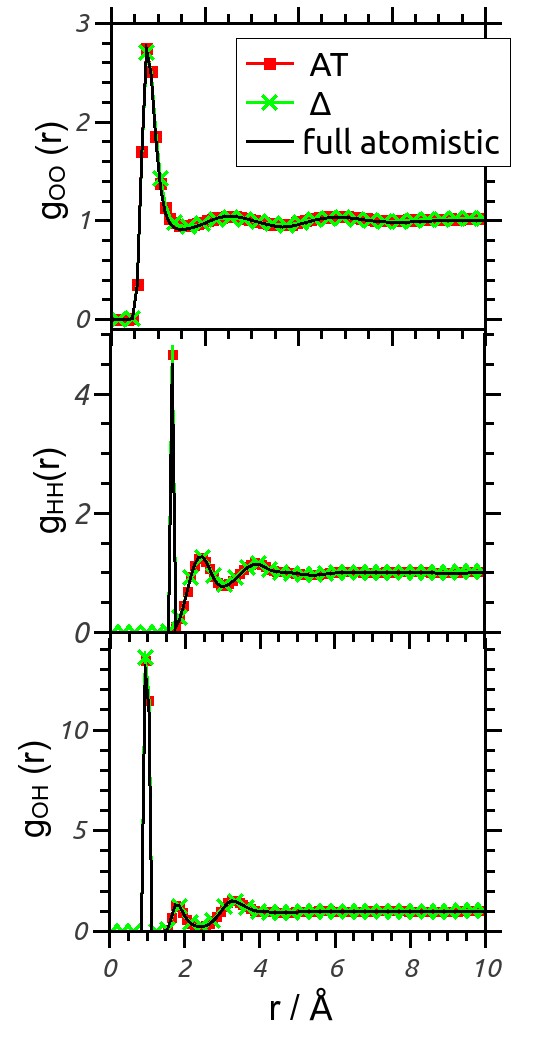}
                \caption{Rdfs: (top) Oxygen-Oxygen,
                  (middle) Hydrogen-Hydrogen, (bottom) Oxygen-Hydrogen. The
                  agreement is highly satisfactory, also in the hybrid region.}
                \label{gofr}
                \end{figure}
Fig.~\ref{gofr} presents the various RDFs
calculated in the AT region, results show a highly satisfactory agreement with reference data. Moreover,
RDFs represent second order (two-body) approximation
of the $N$-particle probability distribution. Hence, the accuracy shown in these
results assures at least a second order accuracy in the statistical
calculation of physical observable in the AT region (see also \cite{prx,njp,krek,shadrack}). The natural question arising is whether such a high accuracy is an
artifact of the scheme, which, by having a direct interface, may actually create
a barrier to the proper exchange of molecules from one region to the
other. In fact, in such a case, the initial configuration (starting from an equilibrated full atomistic setup of the whole simulation box) remains essentially the same in the AT and CG region separately.
 \begin{figure}[htbp]
	        \centering
                \includegraphics[clip=true,trim=0cm 0cm 0cm
                0cm,width=10cm]{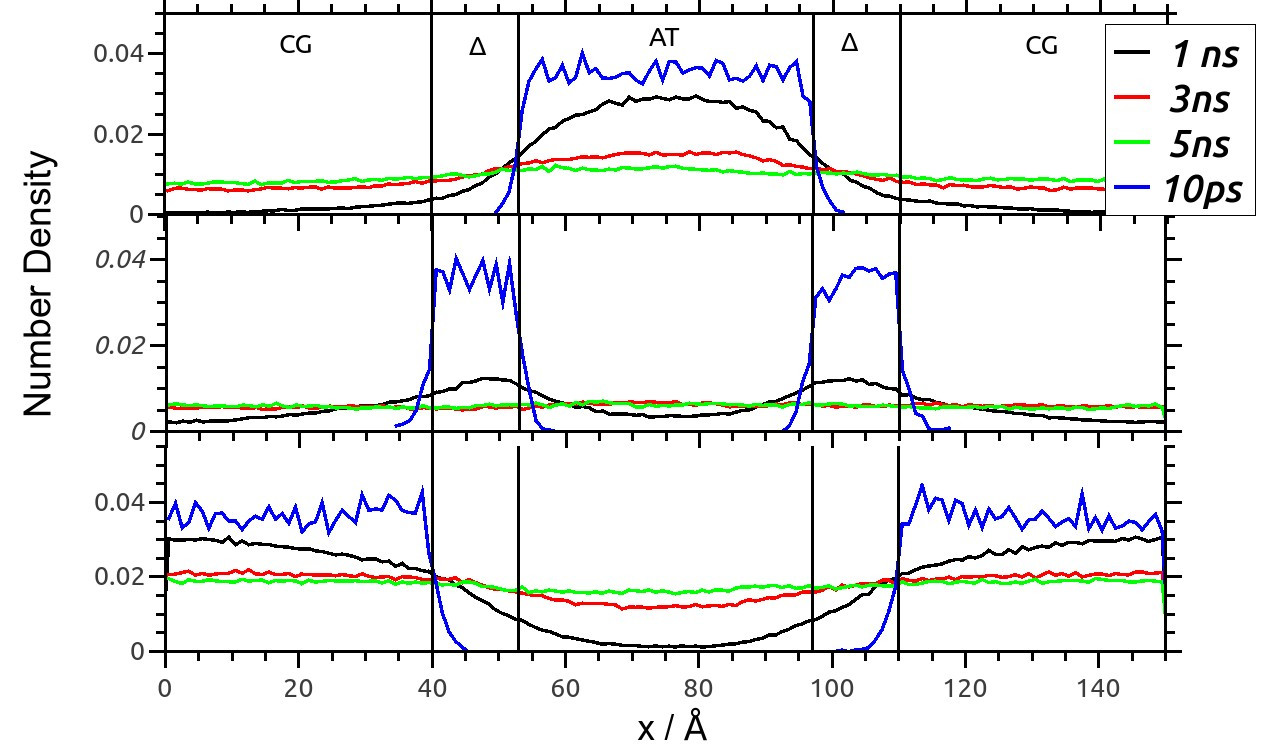}
                \caption{Diffusion of sample of particles taken at $t=0$ in
                  each region and followed in their passage in the other
                  regions as time progresses. The diffusion profile is as expected.}
                \label{diff}
                \end{figure}
Fig.~\ref{diff} shows that samples of particles located in
the AT or CG or $\Delta$ regions properly diffuse into the other regions as the
simulation progresses.  In addition, Fig.\ref{pofn} reports the probability distribution, $P(N)$, in the AT
region and show that it is satisfactorily close (within $2\%$) to the target 
atomistic simulation reference.
\begin{figure}[htbp]
	        \centering
                \includegraphics[clip=true,trim=0cm 0cm 0cm
                0cm,width=8cm]{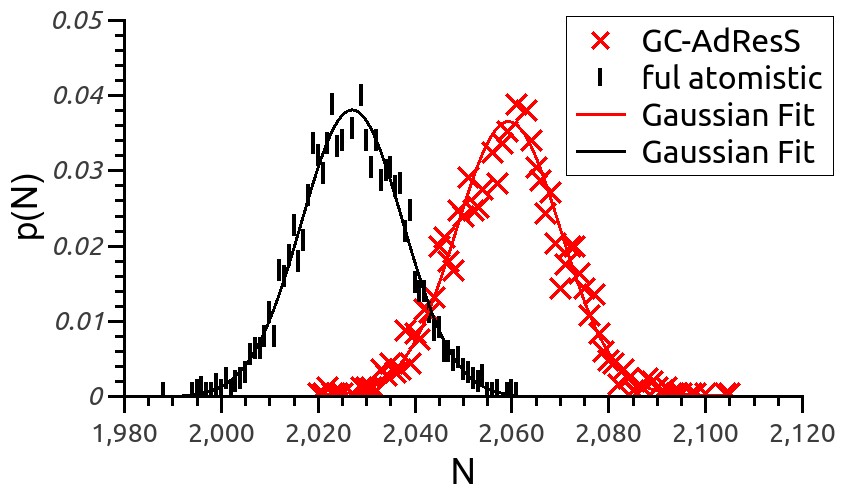}
                \caption{Particle Number probability density in the AT region
                  compared with the equivalent in the full atomistic
                  simulation. The agreement is satisfactory. The $2\%$ deviation is due the accuracy of the thermodynamic force, if desired, this can be systematically improved.}
                \label{pofn}
                \end{figure}
Next, a water molecule has a dipole
whose orientation in the liquid does not have any preferential direction.
However, coarse-grained molecules do not have a dipole and thus one may
envision a sharp artificial order at the interface, typical of polar molecules in contact with a non-polar substrate. 
In turn, this ordering may have strong influence on the liquid structure of the AT region. 
 \begin{figure}[htbp]
	        \centering
                \includegraphics[clip=true,trim=0cm 0cm 0cm
                0cm,width=8cm]{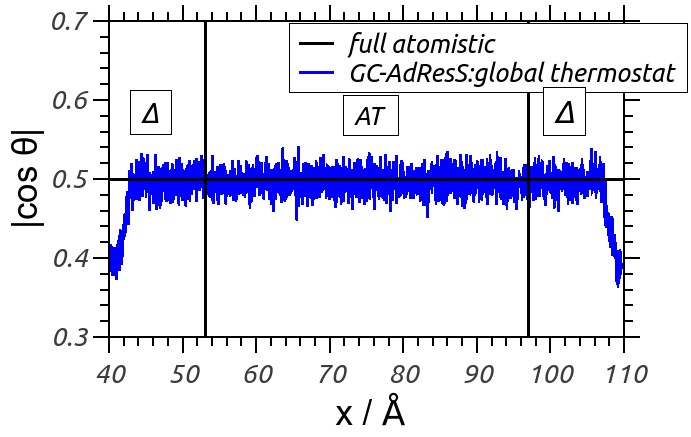}
                \caption{Molecular dipole orientation with respect to the
                  $x$-axis as a function of the position along the region of
                  changing resolution. The agreement with the expected
                  behaviour of a uniform liquid is highly satisfactory. 
                  As expected, at the border between $\Delta$ and the CG region, 
                  the dipole tends to orient due to the non-polar surface of the CG region. 
                  However, such an artificial effect does not have any consequence for the AT region.}
                \label{dip}
                \end{figure}
Fig.~\ref{dip} shows the absence of a preferential alignment of water
dipoles along a direction in the AT region, where in fact a reference
full atomistic simulation is perfectly reproduced. Instead, as expected, at the interface between $\Delta$ and CG regions the molecular dipole tend to align along a preferential direction as it happens to liquid water confined by a non-polar surface. Finally, the
important role of the thermostat may induce doubts about the utility of the
method for the calculation of time-dependent statistical properties since the
dynamics in the AT region would be inevitably perturbed by its action on the
corresponding molecules. However, this problem was already treated in GC-AdResS
by allowing the action of the thermostat in the CG and $\Delta$ regions only, so
that the dynamics in the AT region is the natural dynamics of a Grand
Canonical system \cite{njp,jcppi}. We have implemented the same principle in
the current scheme and confirm its capability to reproduce results of the full
atomistic simulation of reference regarding quantities that characterize the essential features of the liquid. Specifically, the setup with a thermostat acting only in the coarse-grained region shows, as for the setup with a global thermostat reported before, its capability to reproduce (i) the detailed atomistic local structure of the liquid (radial distribution functions, (Fig. \ref{fig1pt})), (ii) the expected exchange of particles in time from one region to another (Fig.\ref{fig2pt}), (iii) the expected particle number distribution in the atomistic region (Fig.\ref{fig3pt}), (iv) the absence in the AT region of artificial orientational effects due to the abrupt change of molecular structure between $\Delta$ and the CG region. In the latter case, to offer an alternative view, we reproduce the tetrahedral order parameter, ${\bf q}$, as a function of the distance (Fig.\ref{tetra}). The latter quantity, for a single molecule $i$ taken as a reference, is defined as: $q_{i}=1-\frac{3}{8}\sum_{j=1}^{3}\sum_{k=j+1}^{4} (\cos \psi_{ijk}+ 1/3)^{2}$, with $j,k$ label the four nearest Oxygen neighbors to the central Oxygen atom of the $i$-th molecule, $\psi_{ijk}$ is the angle 
formed between the vectors {\bf r}$_{ij}$ and {\bf r}$_{ik}$. Finally ${\bf q}$ is calculated as the average over all the water molecule. In addition, we report the dipole-dipole time autocorrelation function, $C_{\mu\mu}(t)$, calculated in the AT region (where there is no thermostat acting) and compared it with the same quantity calculated in the equivalent subregion of the full atomistic (NVE) simulation. The dipole auto correlation function is defined as:
\begin{equation}
C_{\mu\mu}(t) = \frac{1}{N}\sum_{i=1}^{N}\frac{\langle \mu_{i}(t) \cdot \mu_{i}(0) \rangle}{\langle \mu_{i}(0) \cdot \mu_{i}(0) \rangle}
\end{equation}
where  $\langle \mu_{i}(t) \cdot \mu_{i}(0) \rangle$ calculates the correlation between the electric dipole moment of $i^{th}$ molecule at time 0 and $t$. Results show a good agreement with the full atomistic simulation of reference and thus it shows the capability of the method to properly reproduce dynamic quantities as well.
\begin{figure}[htbp]
	        \centering
                \includegraphics[clip=true,trim=0cm 0cm 0cm
                0cm,width=6cm]{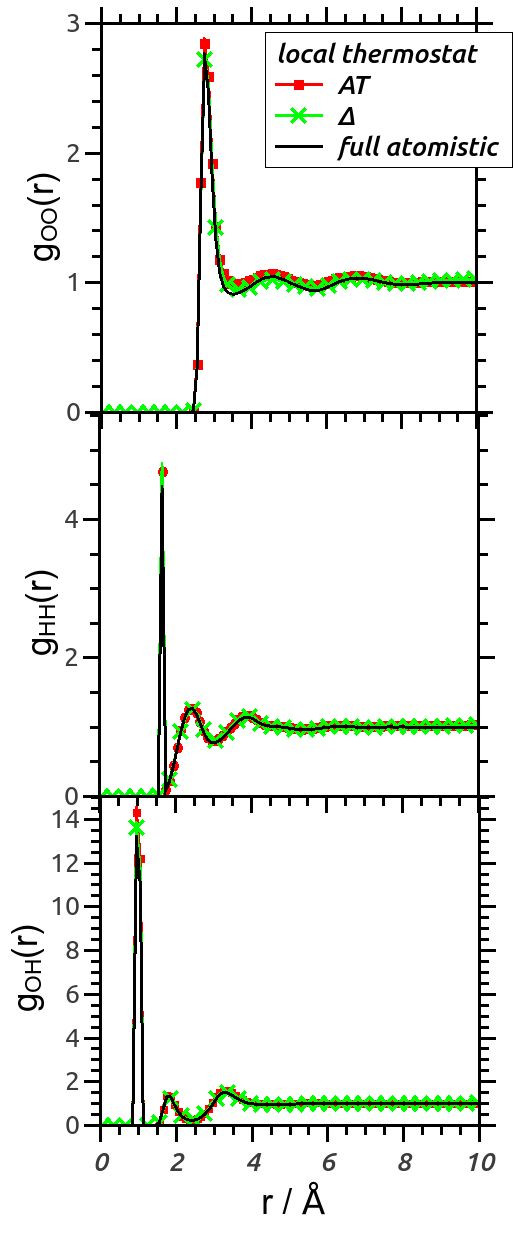}
                \caption{Radial distribution function: (top) Oxygen-Oxygen,
                  (middle) Hydrogen-Hydrogen, (bottom) Oxygen-Hydrogen. The
                  agreement is highly satisfactory.}
                \label{fig1pt}
                \end{figure}
\begin{figure}[htbp]
	        \centering
                \includegraphics[clip=true,trim=0cm 0cm 0cm
                0cm,width=10cm]{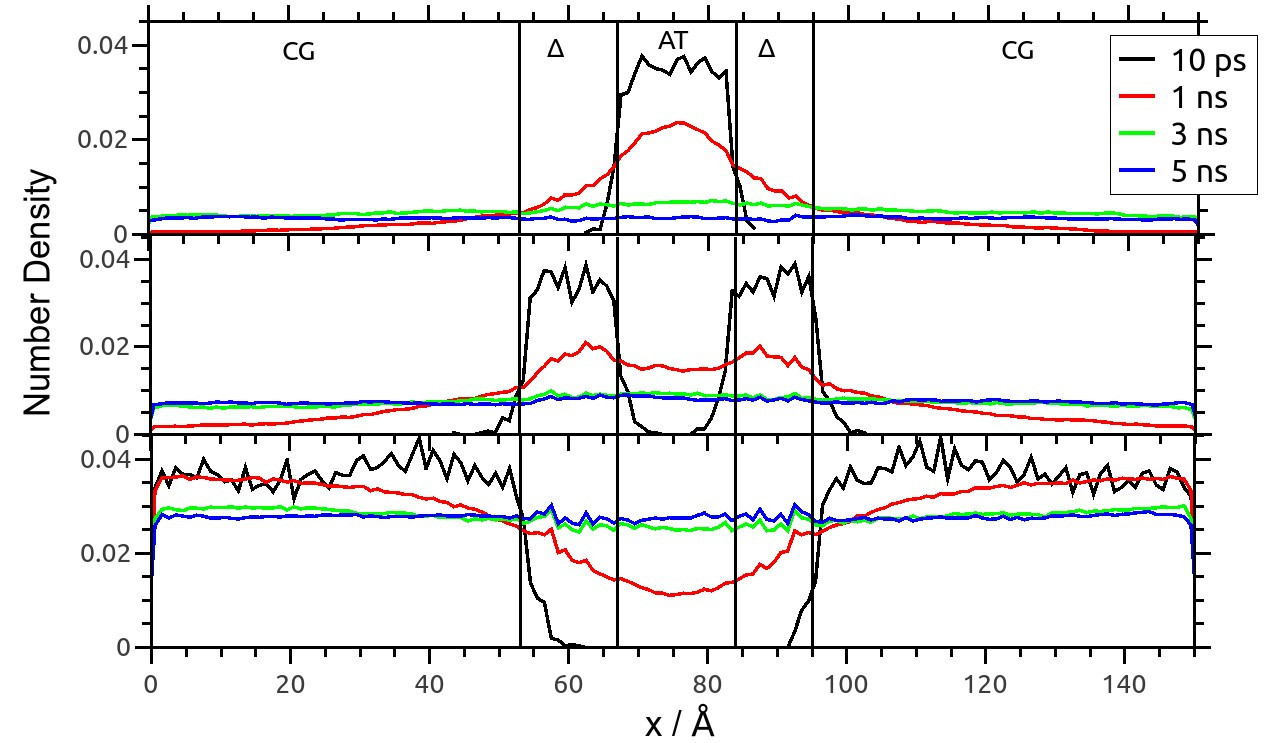}
                \caption{Diffusion of sample of particles taken at $t=0$ in
                  each region and followed in their passage in the other
                  regions as time progresses. The diffusion is as expected.}
                \label{fig2pt}
              \end{figure}
\begin{figure}[htbp]
	        \centering
                \includegraphics[clip=true,trim=0cm 0cm 0cm
                0cm,width=8cm]{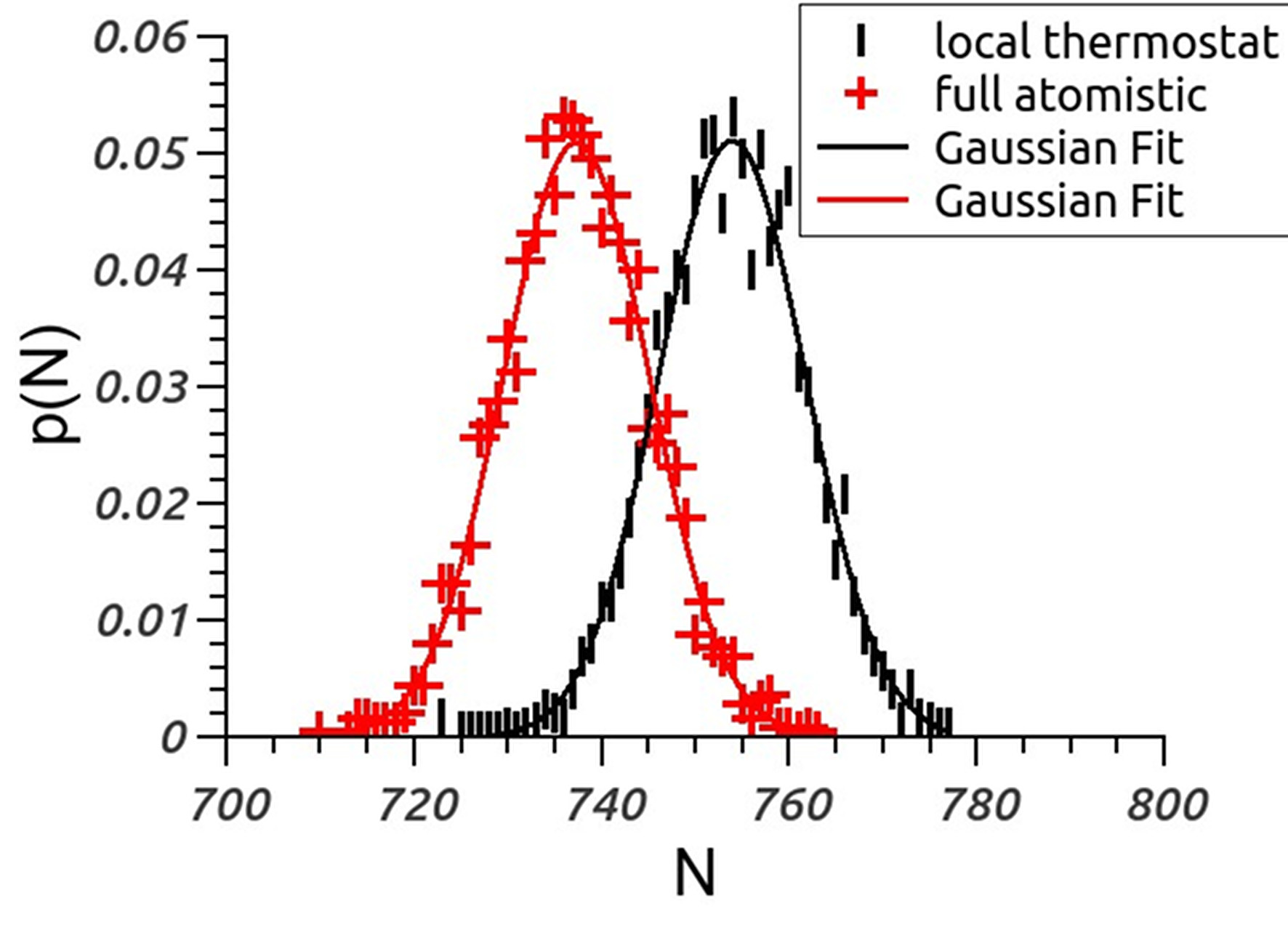}
                \caption{Particle Number probability density in the AT region
                  compared with the equivalent in the full atomistic
                  simulation. The agreement is satisfactory. The $2\%$ deviation is due the accuracy of the thermodynamic force, if desired, this can be systematically improved.}
                \label{fig3pt}
              \end{figure}
\begin{figure}[htbp]
	        \centering
                \includegraphics[clip=true,trim=0cm 0cm 0cm
                0cm,width=9cm]{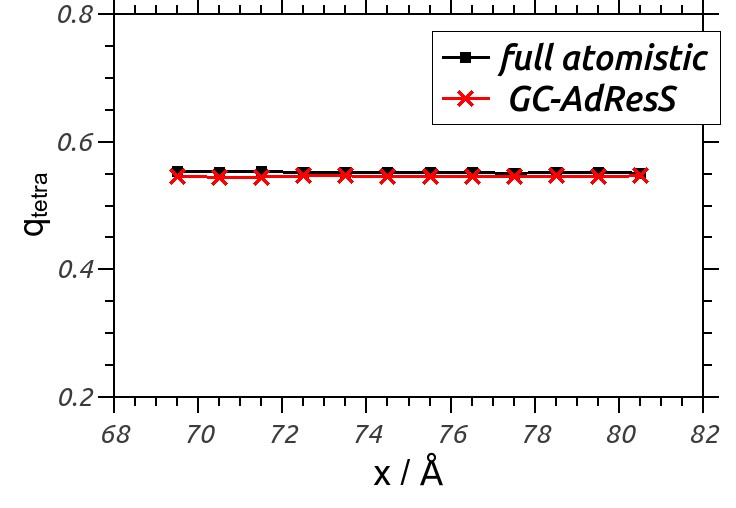}
                \caption{Tetrahedral order parameter of water calculated in the atomistic region of the GC-AdResS simulation with the local thermostat compared to the same quantity calculated in the equivalent subregion in a full atomistic NVE simulation of reference.}
                \label{tetra}
              \end{figure}
\begin{figure}[htbp]
	        \centering
                \includegraphics[clip=true,trim=0cm 0cm 0cm
                0cm,width=8cm]{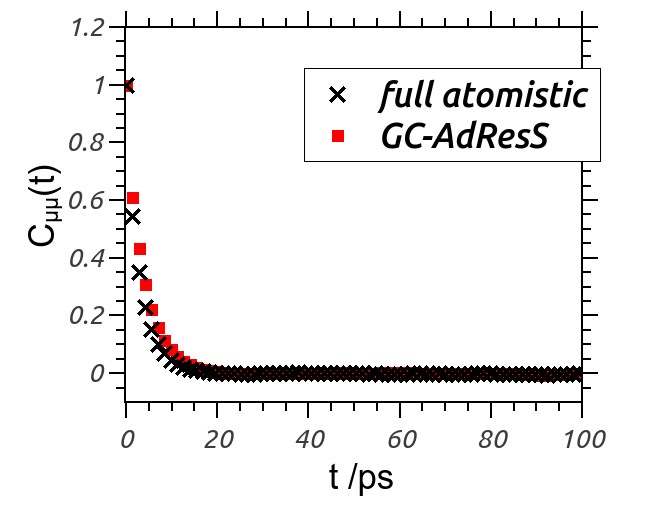}
                \caption{Dipole-dipole time correlation function of water calculated in the atomistic region of the GC-AdResS simulation with the local thermostat compared to the same quantity calculated in the equivalent subregion in a full atomistic NVE simulation of reference.}
                \label{timecorr}
              \end{figure}
Finally,  Fig.~\ref{eff} shows the performance of the current implementation compared to the old
implementation for SPC/E water in the GROMACS code \cite{gromacs}.
 \begin{figure}[htbp]
	        \centering
                \includegraphics[clip=true,trim=0cm 0cm 0cm
                0cm,width=8cm]{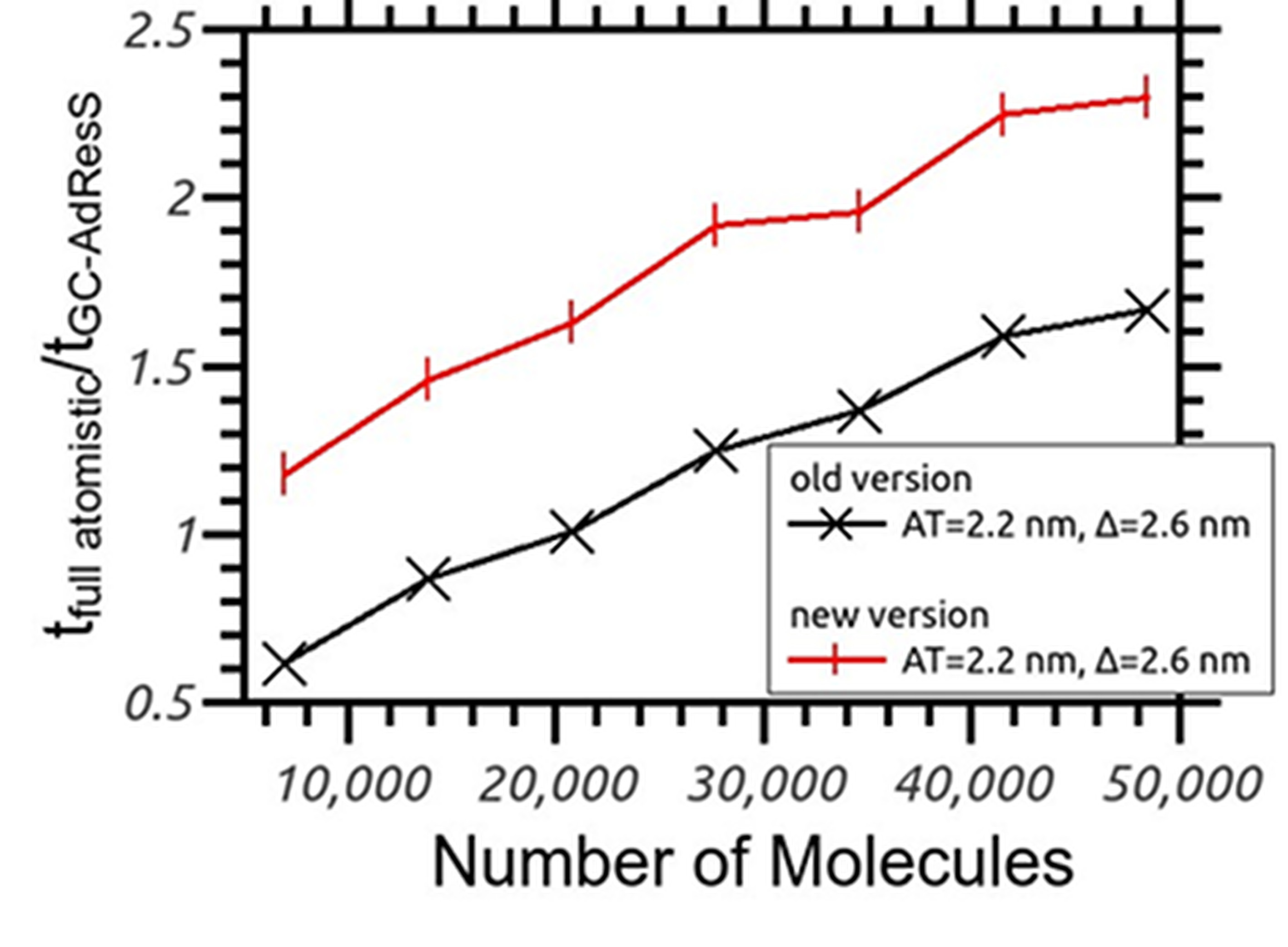}
                \caption{Performance, w.r.t. a full atomistic simulation, of the old version (black) and the new version (red) of GC-AdResS for liquid water at room temperature for the SPC/E water. The AT and $\Delta$ region are kept constant and the CG region is progressively increased.}
                \label{eff}
                \end{figure}
                This is a particularly challenging situation since GROMACS is the most efficient MD code with a specific optimization of the full atomistic simulation of this water model. The enhancement in the performance is sizable (see also the discussion about ionic liquids in the section below).
\subsection{Two micelles in Water}
\begin{figure}[htbp]
	        \centering
                \includegraphics[clip=true,trim=0cm 0cm 0cm
                0cm,width=10cm]{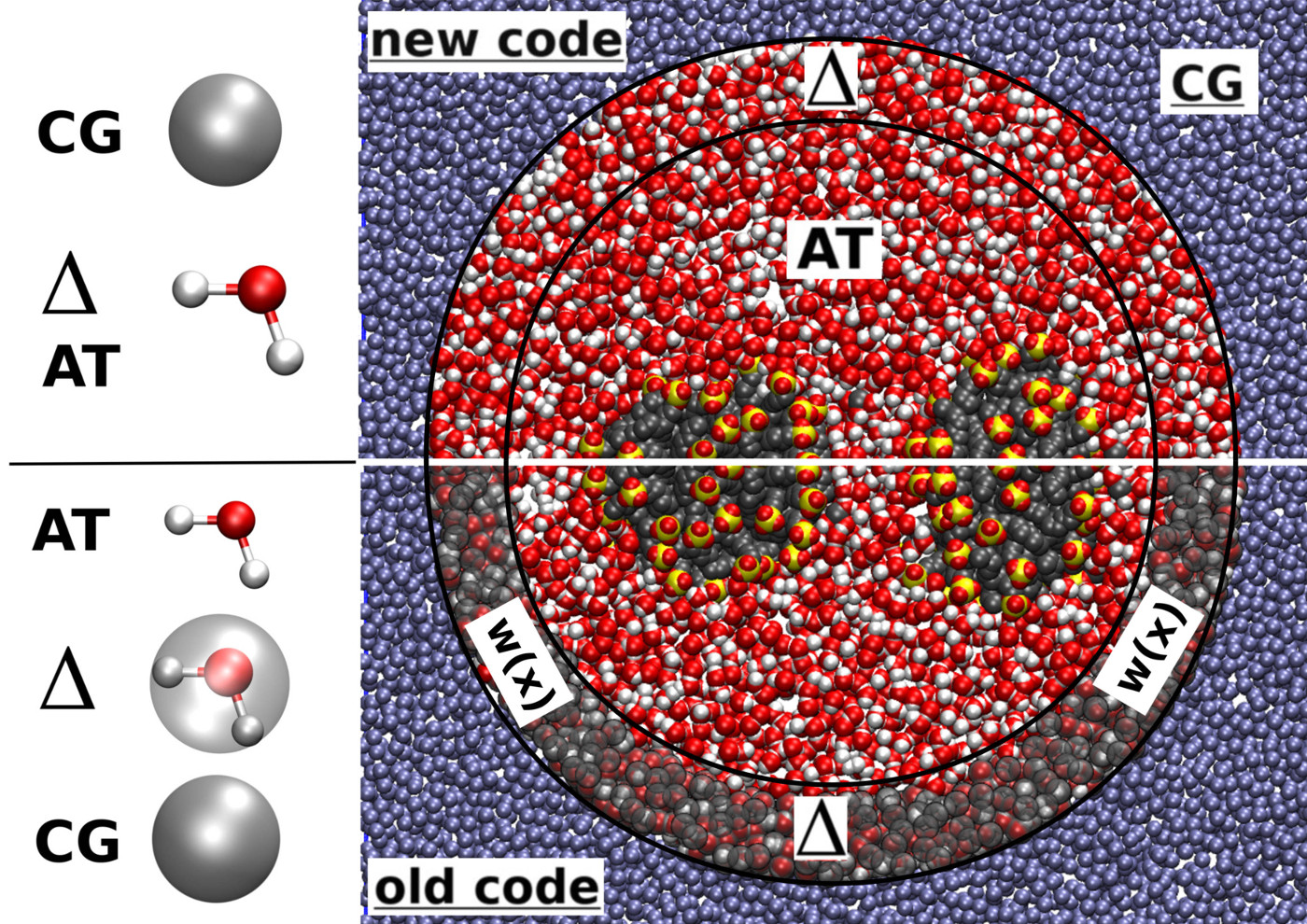}
                \caption{Top: Schematic idea of how the the current version of the model is applied to the solvation of two micelles in water. Bottom: As above, but for the old model with a space-dependent switching function, $w(x)$, in $\Delta$}
                \label{cart-mic}
              \end{figure}

\begin{figure}[htbp]
	        \centering
                \includegraphics[clip=true,trim=0cm 0cm 0cm
                0cm,width=10cm]{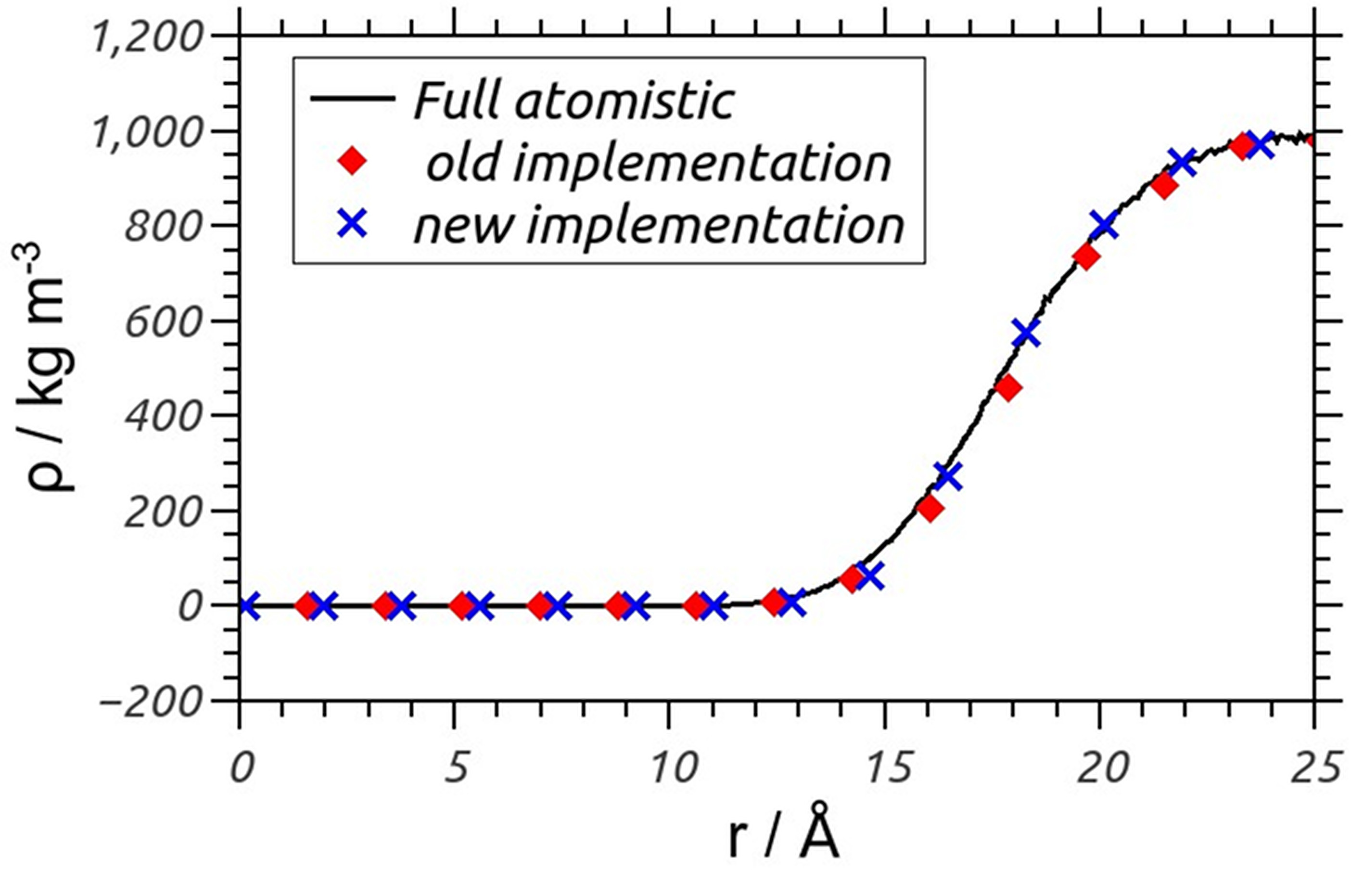}
                \caption{Symmetrized density of water along the center of mass-center of mass direction of the two micelles, analogous to Ref. \cite{shad-adts}.}
                \label{dens-mic}
              \end{figure}

\begin{figure}[htbp]
	        \centering
                \includegraphics[clip=true,trim=0cm 0cm 0cm
                0cm,width=10cm]{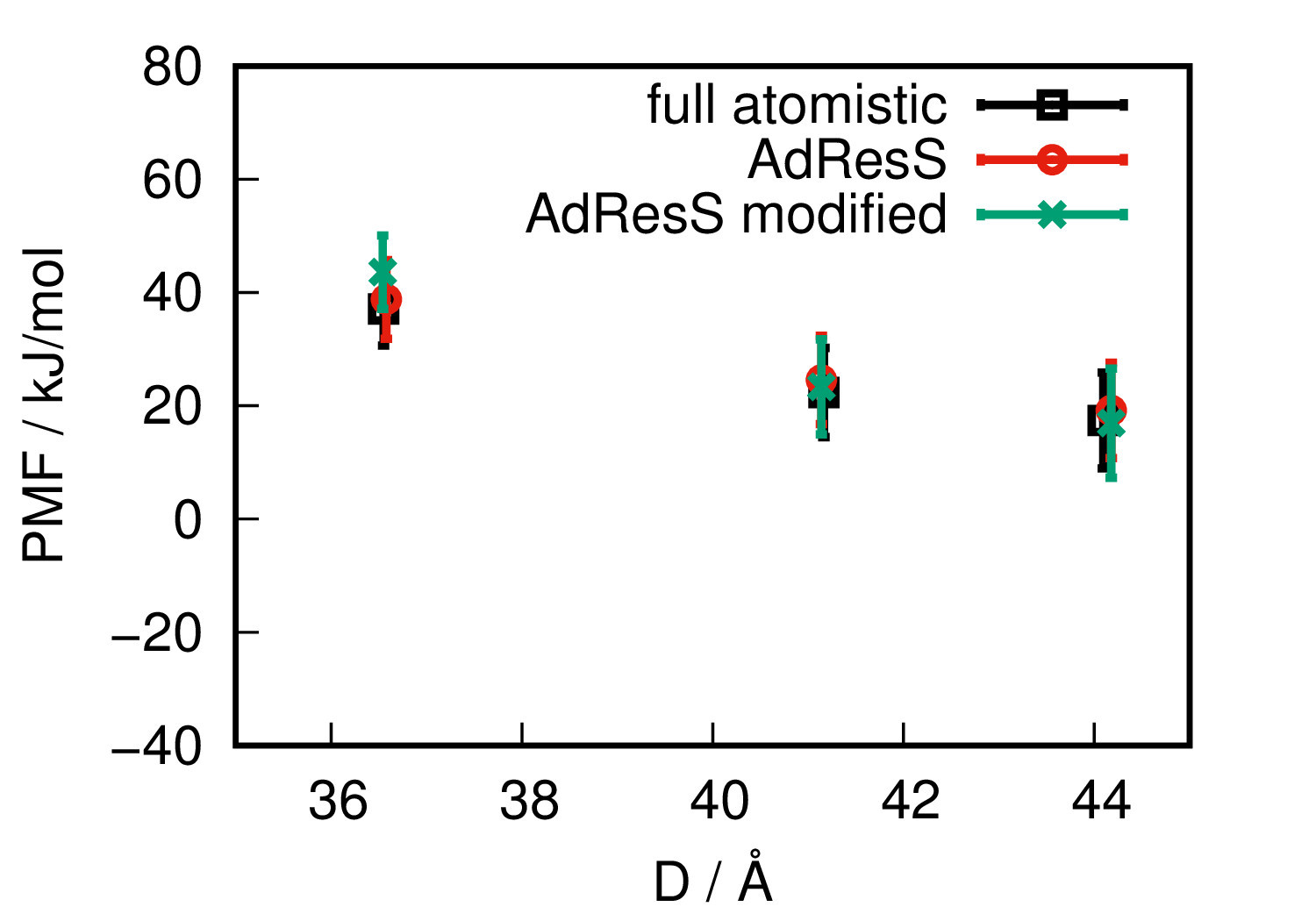}
                \caption{Potential of mean force for three representative micelle-micelle distances from Ref. \cite{shad-adts}, here calculated with the old and new implementation of the AdResS code and compared with the full atomistic simulation of reference.}
                \label{freenergy}
              \end{figure}

  In a recent work by some of us \cite{shad-adts}, GC-AdResS has been employed to determine the free energy of aggregation of two micelles in water.
  This study represents a relevant subject involving a relatively large system and a large number of calculations, thus computational efficiency plays a major role. Fig. \ref{cart-mic} reports a schematic illustration of the two different computational approaches compared here: the current version (top) and the older version of the code (bottom), basic technical details of the simulation can be found in the Appendix, further details can be found in Ref.\cite{shad-adts}.
              Fig.\ref{dens-mic} shows the (symmetrized) average density of water along the center of mass-center of mass direction of the two micelles for a distance of 6.0 nm.
Such a distance was found to correspond to the radius of the smallest atomistic region which can reproduce results of a full atomistic simulation of reference. Results clearly show that the new approach can reproduce the previous AdResS results and the full atomistic data of reference very well. The computational speed up in such case is equal to a factor of 2.4  w.r.t. the full atomistic simulation to be compared with a 1.4 of the old code. Furthermore, in Fig.\ref{freenergy} we plot three relevant points of the free energy of aggregation calculated with the full atomistic approach, with the standard AdResS and with the method proposed in this work. Once again the agreement is highly satisfactory, that is, the most relevant quantity of the study is reproduced with high atomistic accuracy and at the same time at a sizable lower computational price.
\subsection{1,3-dimethylimidazolium chloride Ionic Liquid}
\begin{figure}[htbp]
\centering
\includegraphics[clip=true,trim=0.1cm 0cm 0cm 0.1cm,width=9cm]{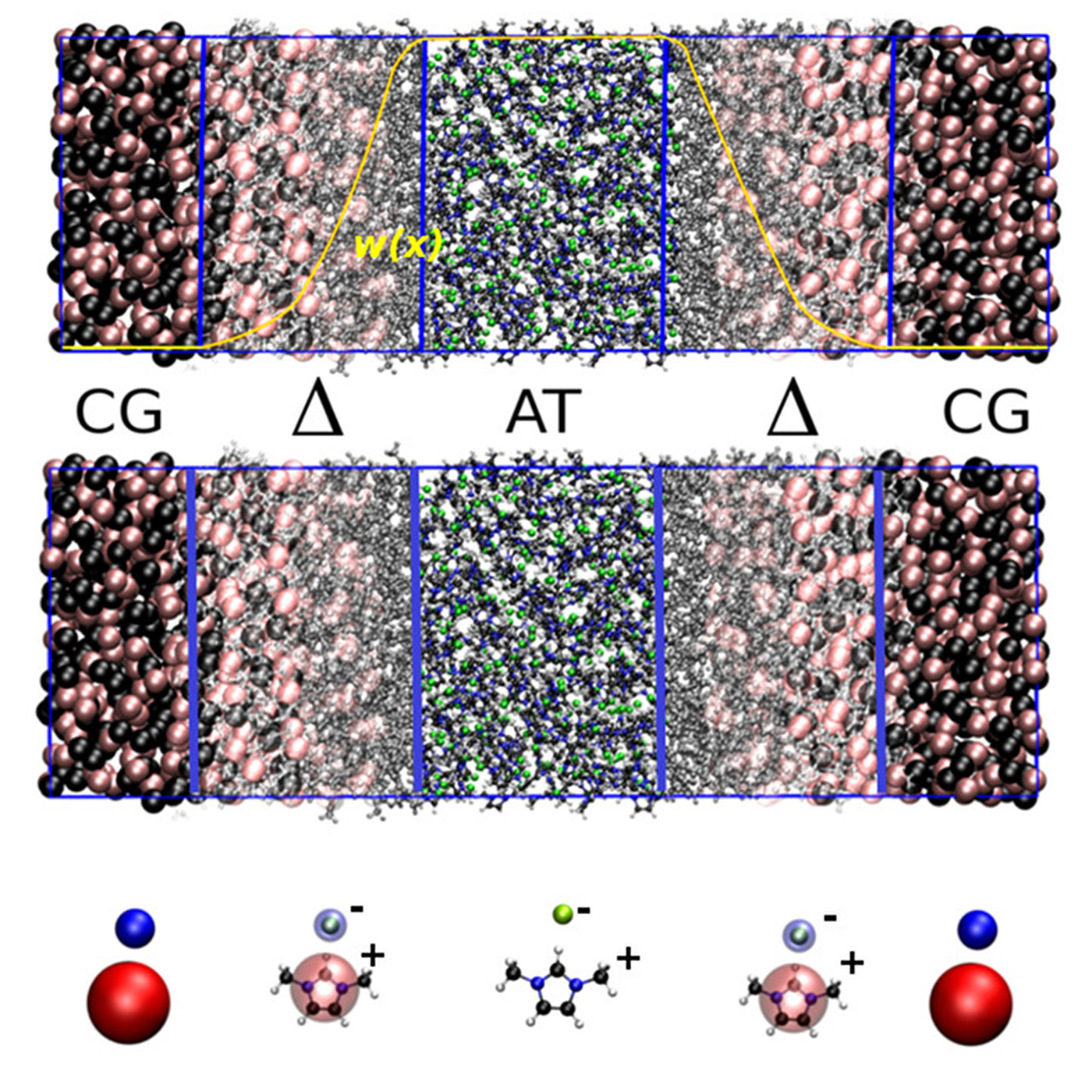}
\caption{(Top) Set-up of standard AdResS for 1,3-dimethylimidazolium chloride Ionic Liquid. The AT region is interfaced with the
  hybrid region $\Delta$, where the molecular
  resolution is weighted by the switching function $w(x)$, and $\Delta$ is interfaced with the coarse-grained region CG.
(Bottom) Set-up in the current work. As for the case of water,in
  $\Delta$, atomistic molecules interact with coarse-grained molecules via
  the coarse-grained potential. We employ a coarse-grained model with uncharged beads, this latter, for the abrupt coupling, is technically far more challenging than the one with charges.}
\label{ilpic}
\end{figure}

\begin{figure}[htbp]
	        \centering
                \includegraphics[clip=true,trim=0cm 0cm 0cm
                0cm,width=8cm]{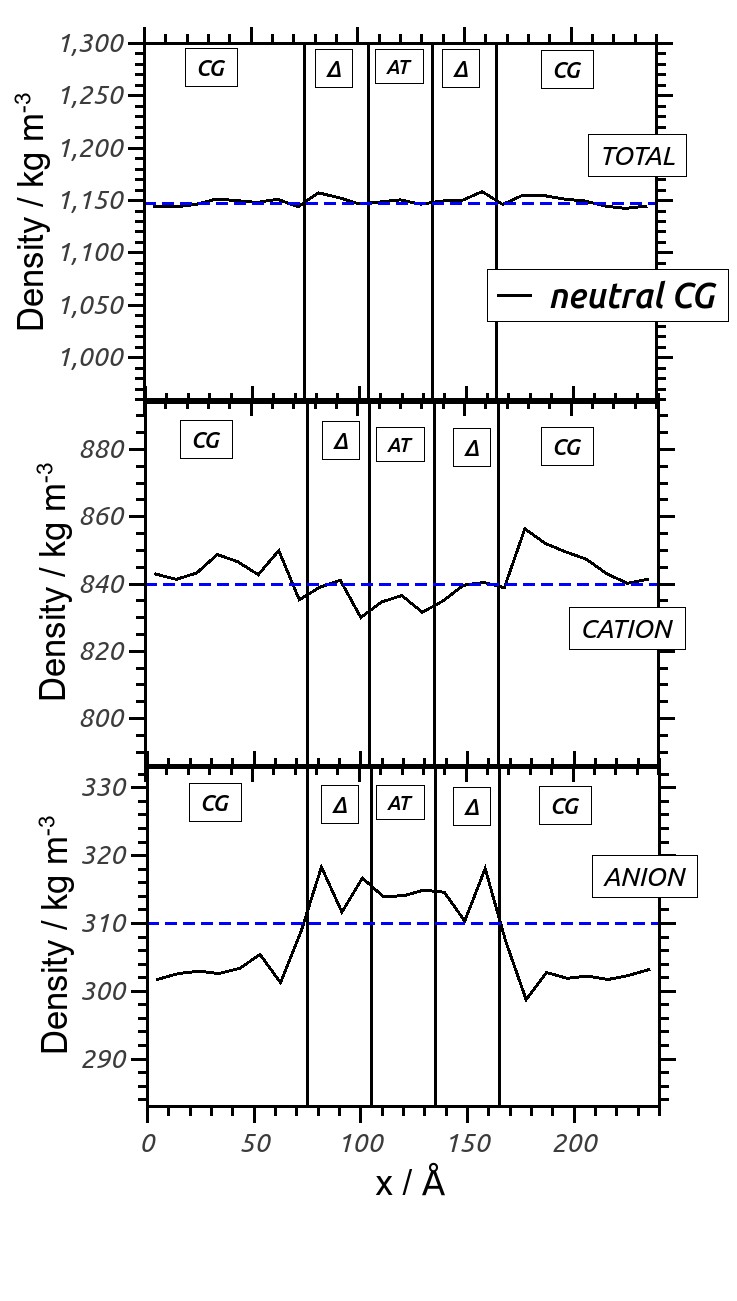}
                \caption{Molecular density of the 1,3-dimethylimidazolium chloride Ionic Liquid in the adaptive box.(top) Density of ions pairs, (middle) density of cations, (bottom) density of anions. All densities in the AT region are within a $3\%$ of difference with respect to the target.}
                \label{densil}
              \end{figure}

\begin{figure}[htbp]
	        \centering
                \includegraphics[clip=true,trim=0cm 0cm 0cm
                0cm,width=8cm]{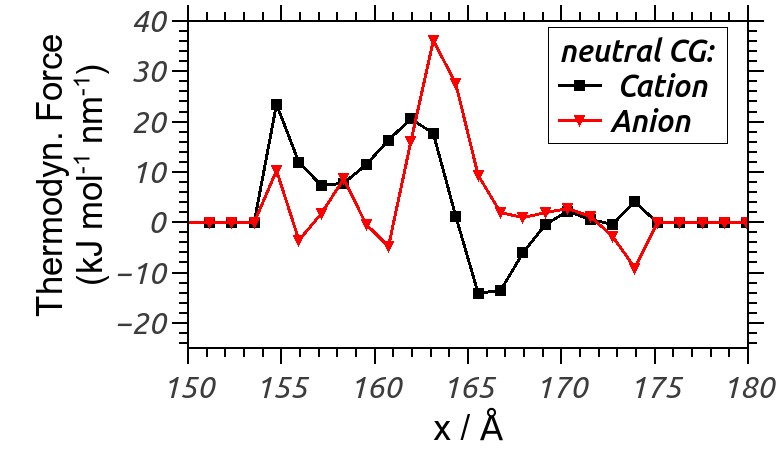}
                \caption{Thermodynamic force as a function of the position in $\Delta$. The black line is the TD for for cations, the red lines for anions. As for the case of liquid water, the action of the force is concentrated at the direct interfacial region. The two thermodynamic forces, as expected, show an (almost exact) antisymmetric behaviour. In fact, the anions being represented by smaller beads diffuse faster towards the interfacial $\Delta$ region and thus tend to have higher density. Instead, the anions being larger diffuse slower and are hindered to occupy the interfacial region in $\delta$ by the high concentration of anions. The thermodynamic force pushes away the anions and drags in the cations in order to establish a density balance of each component.}
                \label{thfil}
              \end{figure}

\begin{figure}[htbp]
	        \centering
                \includegraphics[clip=true,trim=0cm 0cm 0cm
                0cm,width=7cm]{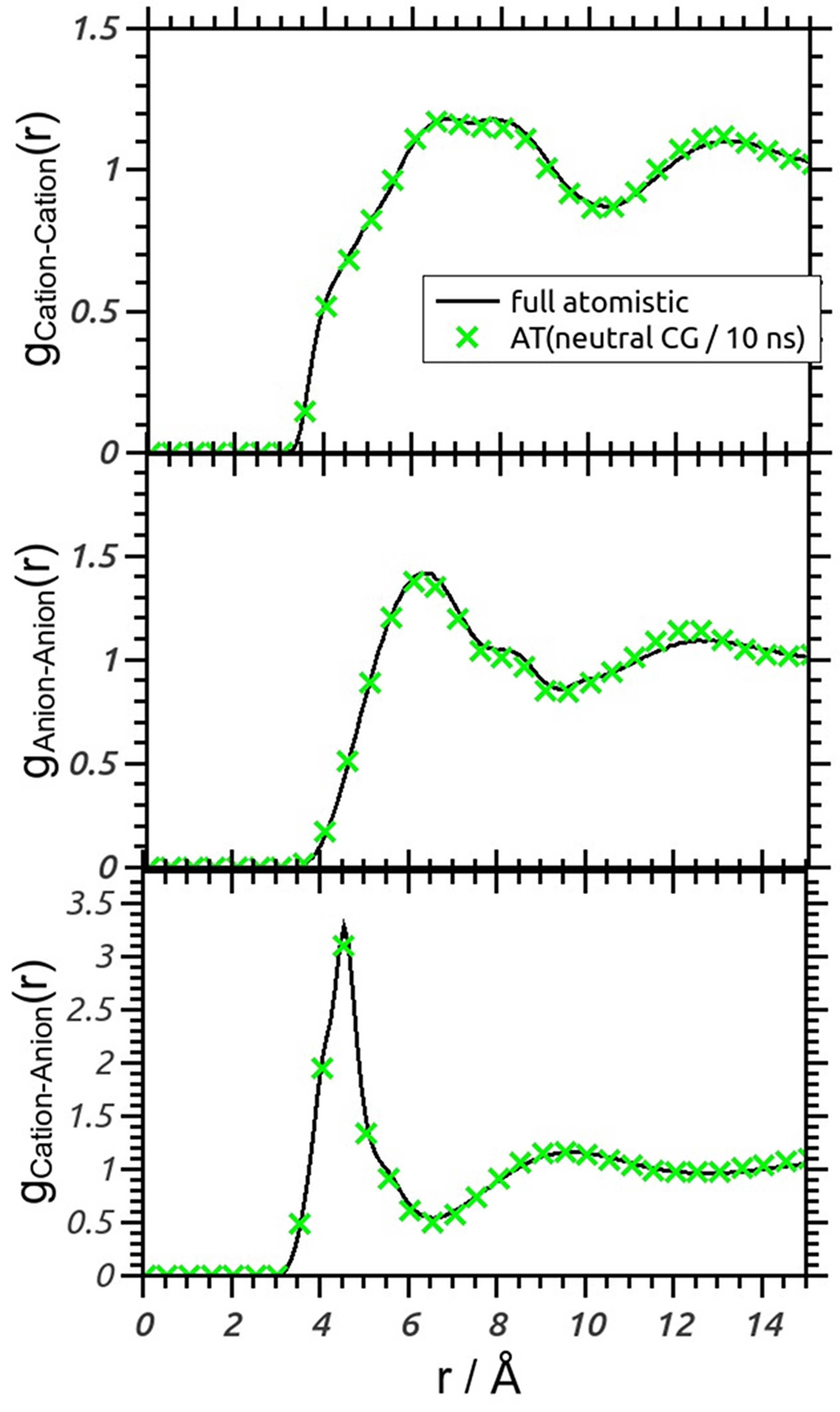}
                \caption{Center of mass-Center of mass radial distribution function in the AT region: (top) Cation-Cation,
                  (middle) Anion-Anion, (bottom) Cation-Anion. The
                  agreement is highly satisfactory.}
                \label{gr1}
              \end{figure}

\begin{figure}[htbp]
	        \centering
                \includegraphics[clip=true,trim=0cm 0cm 0cm
                0cm,width=6cm]{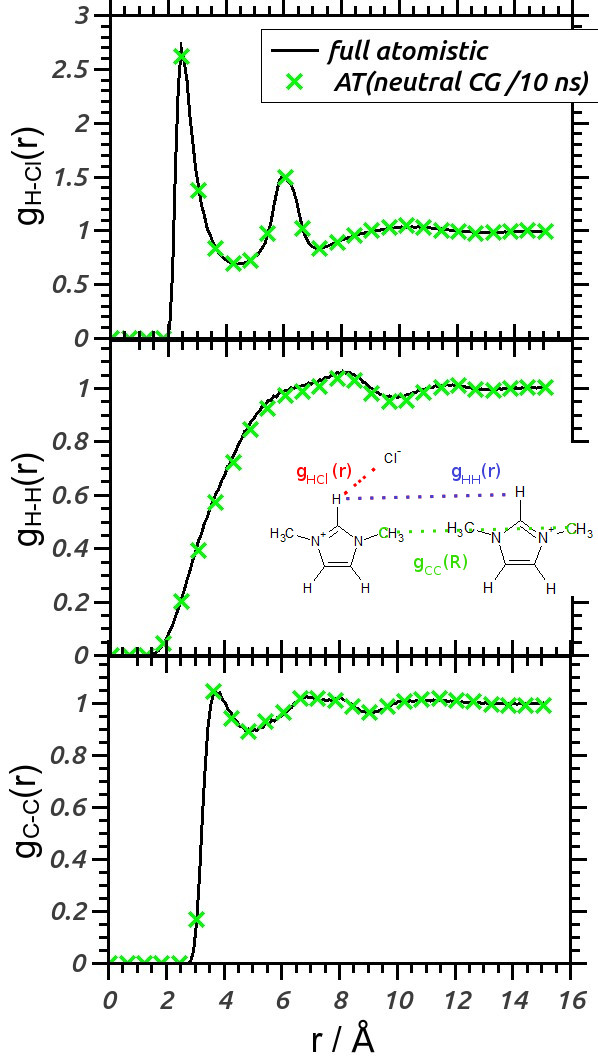}
                \caption{Atom-Atom Radial distribution functions in the AT region according to the definition of the scheme at the top of the figure. The
                  agreement, also in this case, is highly satisfactory.}
                \label{gr2}
                \end{figure}

\begin{figure}[htbp]
	        \centering
                \includegraphics[clip=true,trim=0cm 0cm 0cm
                0cm,width=8cm]{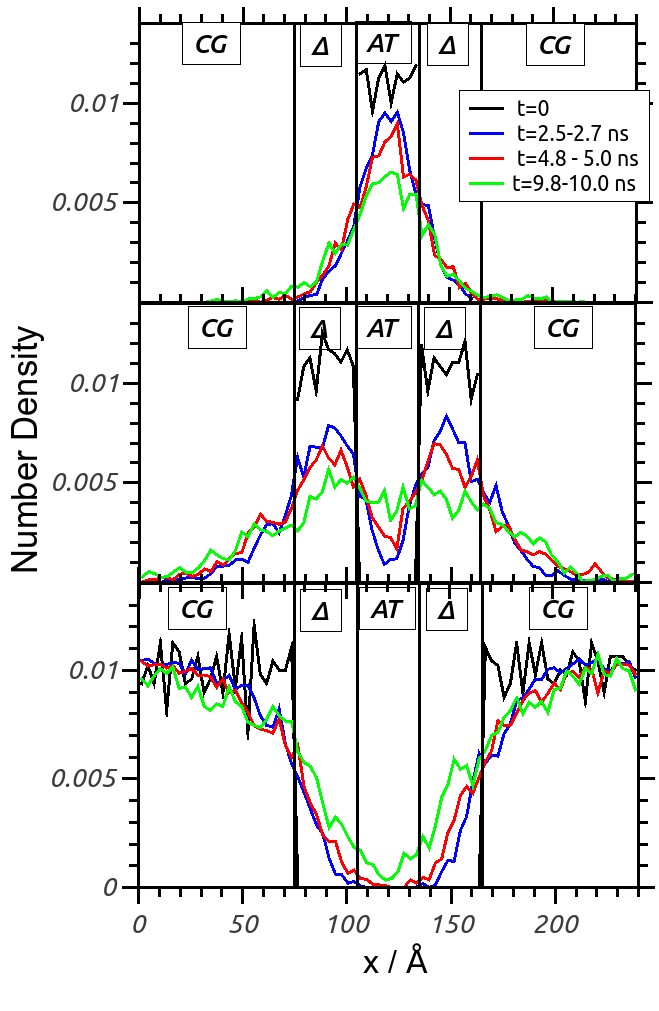}
                \caption{As for liquid water, the diffusion of a sample of particles taken at $t=0$ in
                  each region is followed in the passage to the other
                  regions as time progresses. The diffusion is as expected.}
                \label{diffil}
              \end{figure}

\begin{figure}[htbp]
	        \centering
                \includegraphics[clip=true,trim=0cm 0cm 0cm
                0cm,width=9cm]{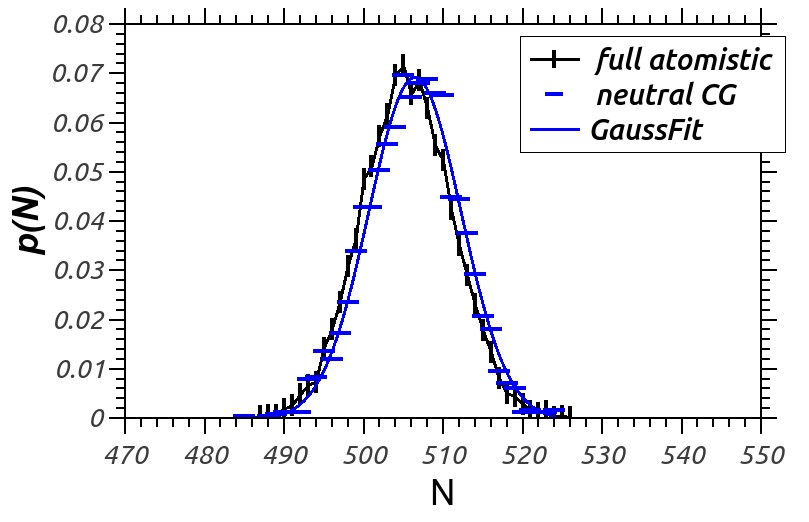}
                \caption{Ion pair number probability density in the AT region
                  compared with the equivalent in the full atomistic
                  simulation. The agreement is satisfactory. As for liquid water, the $1\%$ deviation is due the accuracy of the thermodynamic force.}
                \label{pnil}
              \end{figure}

We have chosen  1,3-dimethylimidazolium chloride ionic liquid as a test system of the new code because this is a typical prototype of ionic liquid and it is highly challenging for MD (in particular for AdResS) given the relevance of the long range electrostatic interactions. The setup of the simulation is the same as that of Ref.\cite{krek} (see Fig.\ref{ilpic}), where the previous version of the code was used.
Further technical details of the simulation are reported in the Appendix.
In particular, we consider an uncharged coarse-grained model, this latter is highly challenging for the abrupt interface employed by the proposed approach.
Similarly to the case of liquid water, we calculate the quantities that characterize the essential features of the liquid and results are reported in Figs. \ref{densil} (see corresponding TD force in Figs.\ref{thfil}), \ref{gr1}, \ref{gr2}, \ref{diffil}, \ref{pnil}. Similarly to the case of liquid water, also in this case the results show that the method has the same accuracy of the previous version of AdResS and an improvement in the computational performance w.r.t. the previous implementation has been found; specifically a speeding factor of about 1.4-1.5 w.r.t. the previous version of the code.
\section{Conclusions}
In conclusion, we have proven that the action of the TD force in
adaptive resolution schemes is technically sufficient for developing a highly efficient and easy-to-implement/trans\-fer\-able algorithm.
Moreover, it assures higher accuracy in the AT region. In future perspective, such a
scheme resolves also the main concern of a recently proposed adaptive
resolution scheme for molecules with electrons \cite{cpcel,adtsluigi}. In fact, in that
scheme, the only technical concern is the absence of a switching function
for a numerical smooth coupling between the quantum and the classical
region. The use of a switching function would make the
quantum treatment artificial and uncontrollably unphysical.
Thus, one can consider the scheme proposed in this work as an alternative path for the construction of an efficient and transferable numerical
recipe for multiscale simulation of last generation.
\section{Appendix: Technical Details}
\subsection{Liquid Water}
As a test system, we simulated a box of water with 6912 SPC/E water molecules and
compared the results of NVT runs with full atomistic MD simulations and also with results from
the standard GC-AdResS simulations. We used a modified GROMACS 5.1.0 \cite{gromacs} code where the box size is a=15.00 nm, b= 3.72071 nm, c= 3.72071 nm. In the
AdResS set-up, the extension of the atomistic region is in total 4.4 nm. The hybrid region
was set to 1.3 nm (a total of 2.6 nm), that is slightly larger than the
cutoff length of 0.9 nm of interactions. The electrostatic interactions are treated by the
standard reaction field method. We use, as in the standard GC-AdResS implementation, the stochastic dynamics integrator (Langevin thermostat). The simulation temperature is 300 K
and the timestep is 2 fs. The thermodynamic force was calculated via 50 ps NVT simulations. Once the thermodynamic force has
converged, we run a 20ns production run. For the case, where a local thermostat was employed, in order to speed up calculations, we kept the
same set-up of the full NVT simulation but we considered a smaller atomistic region (1.6 nm in total). The value of the constant $\kappa$ is equal to 0.07 in all 
  simulations presented in this work, for any system. Our experience suggests that for liquids made of small molecules using this value of $\kappa$
  avoids large jumps in the density as the iterative process evolves and thus leads to a smooth convergence within 25-30 iterations at the worst. It is also 
possible to choose different values and a systematic study of the behavior of the TD as a function of $\kappa$ for different systems and sizes is under development. 
In any case, being an iterative procedure, $\kappa$ does not play a key conceptual role. 
Regarding the simulations used to determine the computational performance of the method w.r.t. to the previous implementations, they were conducted with the modified 
GROMACS 5.1.0 \cite{gromacs} code with systems of 6912, 13824, 20736,
27648, 34560, 41472, 48384 SPC/E water molecules. The total atomistic region was set to 2.2 nm, and the hybrid
region to 1.3 nm (i.e. a total 2.6 nm) while the CG region was systematically increased and run for 5000 steps with a
timestep of 2 fs. The coarse-grained model of water was derived as in Ref.\cite{wat1} by a straightforward Inverse Boltzmann Iterative procedure (IBI) \cite{ibi}. For NVE simulation, we use the reaction field method for calculating the electrostatic interactions in the system,with dielectric constant, $\epsilon_{RF}=\infty$, as this tends to give good energy conservation and has been reported in the gromacs manual. For treating the Vander Waals interactions, we use the ``switch'' method. The cutoff radius for interactions is $0.9$ nm. 

\subsection{Micelles in Water}
We used the same setup as described in Ref. \cite{shad-adts}. The relevant details here are the following: the system consists of a box of $18.0$ nm in linear dimension, containing 190000 water molecules. The atomistic region is a sphere centered in the middle of the box of radius $12.0$ nm containing a hybrid region of a spherical tackiness of $1.25$ nm. For the free energy calculations of the two micelles at the 
different (chosen) points of aggregation we have run simulations  with the above mentioned modified GROMACS 5.1.0 \cite{gromacs} version and compared them with the results of the previous study. All technical details for the reproduction of the results can be found in the technical section of Ref.\cite{shad-adts}.

\subsection{ 1,3-dimethylimidazolium chloride ionic liquid}
We used the same setup for the ionic liquids simulations reported in Ref. \cite{krek}. 
The only difference is that our new system contained 2000 ion pairs of 1,3-Dimethyl-imidazolium 
chloride instead of the published 1000 ion pairs. We duplicated the 1000 ion pairs in x direction. 
We run a 2ns NpT simulation, with T=400K and a Parrinello-Rahman barostat \cite{par_rah}. Then we equilibrated the configuration via NVT for 10 ns at T=400K. The resulting box size was a=23.98304 nm and b=c=3.99718 nm. 
For the GC-AdResS simulations, we used the neutral CG potential already described in Ref. \cite{krek}. 
In the coarse-grained region, the ion pairs are modeled as neutral spherical beads, specifically, one sphere for the cation and one sphere 
with an effective excluded volume, for the anion. The interaction parameters, as for liquid water, are developed using a straightforward 
Inverse Boltzmann Iterative procedure (IBI) \cite{ibi}. Such coarse-grained model have already been tested in several application of AdResS 
\cite{krek,shadrack,krekpshad}. data for analysis are obtained in a production run of 10 ns.



\vspace{6pt} 

\acknowledgments{This research has been funded by Deutsche Forschungsgemeinschaft (DFG) through the grant CRC 1114: ``Scaling Cascades in Complex Systems'', project C01. This work has also received funding from the European Union's Horizon 2020 research and innovation program under the grant agreement No. 676531 (project E-CAM). M. P. acknowledges financial support through the grant P1-0002 from the Slovenian Research Agency.}


\begin{thebibliography}{24}
\expandafter\ifx\csname natexlab\endcsname\relax\def\natexlab#1{#1}\fi
\expandafter\ifx\csname url\endcsname\relax
  \def\url#1{\texttt{#1}}\fi
  \expandafter\ifx\csname urlprefix\endcsname\relax\def\urlprefix{URL }\fi
\bibitem{jcp-adress}
M. Praprotnik, L. Delle Site, and K. Kremer, J. Chem. Phys. {\bf 123}, 224106 (2005)
\bibitem{annurev}
M. Praprotnik, L. Delle Site, and K. Kremer, Annu. Rev. Phys. Chem. {\bf 59}, 545 (2008)
\bibitem{ensing1}
B. Ensing, S. O. Nielsen, P. B. Moore, M. L. Klein, and M. Parrinello, J. Chem.
Theory Comput. {\bf 3}, 1100 (2007)
\bibitem{truhlar}
A. Heyden and D. G. Truhlar, J. Chem. Theory Comput. {\bf 4}, 217 (2008)
\bibitem{csany}
L. Mones, A. Jones, A. W. G\"{o}tz, T. Laino, R. C. Walker, B. Leimkuhler, G. Csanyi and N. Bernstein, J. Comput. Chem., {\bf 36}, 633, (2015)
\bibitem{pre1-adress}
M. Praprotnik, L. Delle Site, and K. Kremer, Phys. Rev. E {\bf 73}, 066701 (2006)
\bibitem{matej-shear}
J. Sablic, M. Praprotnik, and R. Delgado-Buscalioni, Soft Matter {\bf 13}, 4971 (2017)
\bibitem{matej-dna}
J. Zavadlav, R. Podgornik, and M. Praprotnik, Sci. Rep. {\bf 7}, 4775 (2017)
\bibitem{kurt-nucl-ac}
R. Fiorentini, K. Kremer, R. Potestio, and A. C. Fogarty, J. Chem. Phys. {\bf 146}, 244113 (2017)
\bibitem{anim-pccp}
A. Agarwal, C. Clementi, and L. Delle Site, Phys. Chem. Chem. Phys. {\bf 19}, 13030
(2017)
\bibitem{shad-adts}
B. S. Jabes, R. Klein and L. Delle Site, Adv.Th.Sim. 1, 1800025 (2018)
\bibitem{wat1}
M. Praprotnik, S. Matysiak, L. Delle Site, K. Kremer, and C. Clementi, J. Phys.: Condens. Matter {\bf 19}, 292201 (2007)
\bibitem{wat2}
S.Matysiak, C.Clementi, M.Praprotnik, K.Kremer and L.Delle Site, J. Chem. Phys. {\bf 128}, 024503 (2008)
\bibitem{full1}
B. P. Lambeth, C. Junghans, K. Kremer, C. Clementi, and L. Delle Site, J. Chem. Phys. {\bf 133}, 221101, (2010)
\bibitem{prlado}
A. Poma and L. Delle Site, Phys. Rev. Lett. {\bf 104}, 250201 (2010)
\bibitem{prl2012}
S. Fritsch, S. Poblete, C. Junghans, G. Ciccotti, L. Delle Site, and K. Kremer,
Phys. Rev. Lett. {\bf 108}, 170602 (2012)
\bibitem{prx}
H. Wang, C. Hartmann, C. Sch\"{u}tte, and L. Delle Site, Phys. Rev. X {\bf 3}, 011018 (2013)
\bibitem{njp}
  A. Agarwal, J. Zhu, C. Hartmann, H. Wang, and L. Delle Site, New J. Phys. {\bf 17}, 083042 (2015)
\bibitem{obmd}
  R. Delgado-Buscalioni, J. Sablic and M. Praprotnik, Eur. Phys. J. Spec. Top. {\bf 224}, 2331 (2015)
\bibitem{obmd-DNA}
  J. Zavadlav, J. Sablic, R. Podgornik and M. Praprotnik, Biophys J. {\bf 114}, 2352 (2018)
\bibitem{prlraff1}
R. Potestio, S. Fritsch, P. Espanol, R. Delgado-Buscalioni, K. Kremer, R. Everaers, and D. Donadio, Phys. Rev. Lett. {\bf 110}, 108301 (2013)
\bibitem{prlraff2}
R. Potestio, P. Espanol, R. Delgado-Buscalioni, R. Everaers, K. Kremer,
and D. Donadio, Phys. Rev. Lett. {\bf 111}, 060601 (2013)
\bibitem{jcp-simon}
S. Poblete, M. Praprotnik, K. Kremer, and L. Delle Site, J. Chem. Phys. {\bf 132}, 114101 (2010)
\bibitem{jctc-han}
H. Wang, C. Sch\"{u}tte, and L. Delle Site, J. Chem. Theory Comput. {\bf 8}, 2878 (2012)
\bibitem{jcp-mu}
  A. Agarwal, H. Wang, C. Sch\"{u}tte, and L. Delle Site, J. Chem. Phys. {\bf 141}, 034102 (2014)
\bibitem{pre2007}
L. Delle Site, Phys.Rev.E, {\bf 76}, 047701 (2007)
\bibitem{jcppep}
P. Espanol, R. Delgado-Buscalioni, R. Everaers, R. Potestio, D. Donadio
and K. Kremer, J. Chem. Phys. {\bf 142}, 064115 (2015)
\bibitem{physrep}
L. Delle Site and M. Praprotnik, Phys.Rep. {\bf 693}, 1-56 (2017)
\bibitem{krek}
C. Krekeler and L. Delle Site, Phys. Chem. Chem. Phys. {\bf 19}, 4701 (2017)
\bibitem{shadrack}
  B. S. Jabes, C. Krekeler, R. Klein and L. Delle Site, J. Chem. Phys. {\bf 148}, 193804 (2018)
\bibitem{krekpshad}
 B. S. Jabes and C. Krekeler, Computation, {\bf 6}, 21 (2018) 
\bibitem{hmmm}
  D. de las Heras and M.Schmidt, https://arxiv.org/abs/1711.00006
\bibitem{jan}
  J. H. Peters, R. Klein and L. Delle Site, Phys.Rev.E {\bf 94}, 023309 (2016)
\bibitem{neumann}
  M. Neumann, J. Chem. Phys. {\bf 82}, 5663, (1985)
\bibitem{amdahl}
  C. Junghans, A. Agarwal and L. Delle Site, Comp. Phys. Comm. {\bf 215}, 20 (2017)
  \bibitem{cameron}
 C.F. Abrams, J.Chem.Phys. {\bf 123}, 234101 (2005)
\bibitem{jcppi}
  A. Agarwal and L. Delle Site, J. Chem. Phys. {\bf 143}, 094102 (2015); ibid Comp. Phys. Comm. {\bf 206}, 26 (2016)
\bibitem{gromacs}
M. J. Abraham, T. Murtola, R. Schulz, S. Pall, J. C. Smith, B. Hess, E. Lindahl, Software X, {\bf 1-2}, 19, (2015)
\bibitem{cpcel}
  L. Delle Site, Comp. Phys. Comm. {\bf 222}, 94 (2018)
\bibitem{adtsluigi}
L.Delle Site, Adv.Th.Sim. (2018), in press
\bibitem{ibi}
D.Reith, M.P\"{u}tz and F M\"{u}ller-Plathe, J. Comp. Chem. {\bf 24}, 1624 (2003)
\bibitem{chemphyschemrev}
  F.Dommert, K.Wendler, R.Berger, L.Delle Site and C.Holm, ChemPhysChem {\bf 13}, 1625 (2012)
\bibitem{berendsen}
  H.J.C. Berendsen, J.P.M. Postma, W.F. van Gunsteren, A. DiNola, J.R. Haak, J. Chem. Phys. {\bf 81}, 3684 (1984)
\bibitem{par_rah}
  M. Parrinello, A. Rahman, J. Appl. Phys. {\bf 52}, 7182 (1981)
\bibitem{jcp_exp}
C.Hardacre, J.D.Holbrey, S.E.J.McMath, D.T.Bowron, and A.K.Soper, J. Chem. Phys. 118, 273 (2003)
\end{thebibliography}
\end{document}